\documentstyle[11pt]{article}
\newcommand{\blankline}{\vskip .3cm}
\newcommand{\f}{\begin{equation}}
\newcommand{\ff}{\end{equation}}

\begin{document}
\centerline{\LARGE Coevolution of membranes and channels:}
\centerline{\LARGE A possible step in the origin of life}
\blankline
\rm
\centerline{Saint Clair Cemin and Lee Smolin${}^*$}
\blankline
\centerline{\it  ${}^*$ Center for Gravitational Physics and Geometry}
\centerline{\it Department of Physics}
 \centerline {\it The Pennsylvania State University}
\centerline{\it University Park, PA, USA 16802}
\blankline
\blankline
\blankline
\blankline
\centerline{September 22, 1997}
\blankline
\blankline
\blankline
\blankline
\centerline{ABSTRACT}

We propose a scenario for the origin of life based on the
coevolution of lipid bilayer vesicles and protein
 channels.

\blankline
\vfill
${}^*$ smolin@phys.psu.edu
\eject

\section{Introduction: Assumptions}

We propose here an idea concerning the origin of life which
we believe to be novel, but closely related to some previous ideas
on the subject\cite{harold1,lipids}.   
The basic idea is that the first self-reproducing
entities, autonomous agents in Kauffman's sense\cite{stu-agents} 
were lipid
vesicles with simple protein channels.  We will argue that there
is a plausible sequence of events whereby such entities could
have arisen from lipids and amino acids present in some
primordial body of water, and co-evolved to become a primitive
form of stable, self-reproducing entities.

We begin by listing the assumptions on which our proposal is
based.

\begin{itemize}

\item{}The first self-reproducing entities arose in solution
in lakes, oceans or pools of water on the early earth.  They
arose from interactions of small molecules formed by the
passage of energy through the medium, by the 
Miller-Urey synthesis
or some analogue\cite{miller}.    
The energy may have come from heat,
lightning, reaction of redox couples or solar photons.  

\item{}Experiment has shown that the passage of energy
through a medium of simple molecules such as
$CO_2, NH_3,CH_4$ and $H_20$ may result
in the formation of amino acids\cite{miller}.  
Lipids and amphiphilic molecules are also observed to
form under similar primitive conditions\cite{bilayer,harold2}.
Once formed they spontaneously
form bilayer membranes whose minimal energy configurations are hollow
vesicles\cite{bilayer,harold2}.  Such vesicles 
have even been seen to
form spontaneously in material dissolved from
meteorites\cite{cosmolipids}.  

\item{}At the same time the 
evidence seems to indicate that  nucleic acids do not
form as readily or abundantly in such simple 
processes\cite{nonucleic}.  Although
the hypothesis that the first living things were RNA
molecules has been widely considered, 
there has yet to be discovered a convincing
scenario whereby the ingredients for RNA could have
formed in solution from prebiotic materials\cite{nonucleic}.  
Nor is there yet a convincing scenario for the assembly of
RNA or DNA spontaneously from its components, in the
absence of the action of 
enzymes\cite{nonucleic}\footnote{There are interesting
suggestions about the role of clay\cite{clay} in the origin
of life, but these suffer from an important problem which
a vesicle first scenario does not, which is how a network of
chemical reactions that develops using clay as a kind of
enzyme should transmute itself to one that has no trace
of a role for clay or its constituents in it\cite{harold2}.}.

Nucleic acids involve a higher stage of chemical 
complexity\cite{harold2,harold-sfi}.  They also require special
conditions present within cell membranes and absent in
sea water, such as the absence of calcium, which degrades
them.  Therefor it seems unlikely that nucleic acids evolved
until after there were primitive vesicles with a mechanism for
excluding calcium.

\item{}It seems most probable that the   
the first self-reproducing
entities formed from the most common materials
out of which they may be constituted. It therefore seems
rational to ask if there is a possible scenario by which 
amino acids and lipids might have constituted the first
self-reproducing entities.  If so, given the evidence that they
form more easily and copiously than nucleic acids 
in plausible prebiotic environment, it seems worthwhile to investigate
the hypothesis that the first self-reproducing entities were
composed solely of lipids and amino acids.

\item{}It is known experimentally that polar 
lipids will spontaneously
form vesicles formed from bilayers 
in solution\cite{bilayer,harold2}.  These may spontaneously grow
by the accumulation of lipid molecules and they have been
observed to spontaneously divide\cite{bilayer}.  It thus seems most
parsimonious to assume that the first self-reproducing
entities involved these lipid bilayer 
vesicles\footnote{The suggestion that the origin of
life began with the spontaneous formation of lipid vesicles
is of course very old.  See, for 
example\cite{harold1,harold2,harold-sfi} and references
contained therein.}.  Given that they
both form spontaneously in models of primordial environments
and are in fact essential components of living cells it would
be strange if they were not part of the first self-reproducing
entities.

\item{}At some early stage some primitive metabolism must
spontaneously begin and it must be coupled to the early
self-reproducing entities.  It is natural to search for the
origins of such processes in the phase separation accorded by
the bilayer lipid vesicles.  We will not discuss here the
possible nature of the primitive metabolism, except to assume
that it begins spontaneously within vesicles 
when they are illuminated by light.  All that is needed is
that light be converted to some form of potential energy by
bonding or dissasociating small molecules. The vesicle plays
a role to contain the molecules involved, and prevent their
dillution.  Plausible scenarios for how this may have occurred,
for example from the presence of chromophores dissolved
in the interiors of the membrane vesicles, are described
in \cite{harold2,harold-sfi}.

There is, however, a serious problem
with the hypothesis that the first self-reproducing entities
involved only lipid vesicles coupled to some simple metabolism.
This is the osmotic crisis\cite{osmotic}.  
The problem arises because, as just
mentioned, simple metabolic
processes will be more likely to occur in the interior of lipid
vesicles, as the concentration of their products will not be diluted
by diffusion, because they cannot pass through the membrane.
On the other hand, for exactly the same reason, any concentration
difference in a molecule or ion incapable of diffusing through
the membrane leads to a buildup of an osmotic pressure across
the membrane.   If not somehow vented, the pressure will
rupture the membranes.   Thus, it seems that there is a kind of
paradox concerning the role of membranes in the origin of
metabolism.  Membranes may have played an essential role
by concentrating the
products of metabolism so that some autocatalytic cycles may
establish themselves.  At the same time, any sufficiently active
metabolism results in the buildup of osmotic pressure which
ruptures the membranes.  We will refer to this as the
{\it osmotic paradox.}

\item{}Since amino acids also form spontaneously by passage
of energy through plausible primordial media, it is natural
to look for a role of proteins also in the first self-reproducing
entities.  Here, however, we meet a problem, which is that
amino acids do not appear to spontaneously form themselves
into proteins.  The reasons for this are two-fold: First the
binding of amino acids into proteins produces water, and
this is not favored in an aqueous environment.  Second,
the rates at which amino acids at relatively low concentrations
in solution meet each other are not very high.

There may be some spontaneous formation of short polypeptides
through processes such as repeated dehydration. These may
have occurred in puddles of water, repeatedly filled by
rainwater and evaporated by sunlight.  The problem is that
the periodic evaporation poses a problem for the survival
of any entities formed in the puddle.  And if, on the other hand,
the contents of such puddles were from time to time washed into
a lake or ocean, there would be the immediate problem of the
dillusion of the polypeptides.  

\end{itemize}

\section{Interactions of amino acids and lipid membranes}

Given the assumptions we have just listed, it is natural to look
for an origin of the first self-reproducing entities in some interaction
of amino acids and lipid membranes.  It seems quite plausible that
both were found to significant concentration in some primitive
aqueous environments.  However, by themselves, each component
has a problem that seems to block its alone constituting the
first self-reproducing entity.  In the case of the lipid membranes
it is the osmotic paradox we described.  In the case of the amino
acids it is the fact that they do not spontaneously form into
proteins.

Living things, by definitions, are entities that ensure their
survival by solving whatever problems they are faced with.
(One formalization of this statement is Kauffman's characterization
of an autonomous agent\cite{stu-agents}.)
It is then natural if the origin of life involves the discovery
of the solution to a small set of problems that block the
synthesis or survival of some set of molecules.  
Given that both lipid membranes and amino acids each
face such a problem, which prevents them from individually
constituting the first self-reproducing entities, in spite of their
being common in plausible primitive environments, we may
ask if there are mechanisms by which they could solve each
others problem.  In this case the first self-reproducing entities
might arise by coevolution of the lipid boundaries and proteins.

Once this question is stated some natural hypotheses suggest 
themselves.  One particularly attractive set are the following.

\subsection{How the membranes may solve the proteins' problem}

\begin{itemize}

\item{}{\bf H1}  {\it In an aqueous medium containing 
lipid bilayer vesicles and amino acids, concentrations
of amino acids will develop 
on the outer surfaces of lipid vesicles.}
Some amino acids have hydrophobic groups, while others
have both hydrophobic and hydrophilic ends.  The hydrophobic
groups will be attracted to the lipid-water boundary.
The hydrophobic amino acids may also bury themselves inside
the lipid membranes.  The result may be a steady buildup
of amino acids on the outer surface of the vesicles.

\item{}{\bf H2} {\it Once present on the surface of the vesicle, amino
acids will react to form polypeptides. } These polypeptides will be
randomly formed from the species of amino acids that
develop concentrations on the surfaces of the vesicles.
The reason for this is that the lipid-water boundary may solve
both problems that block the binding of amino acids into
proteins in the aqueous medium. First, amino acids and
polypeptides are far more likely to encounter each other when
they are restricted to move on a two dimensional surface than
in solution. The finite size of the vesicles increases this probability
further, for as long as the vesicle survives it can accumulate
an increasing surface density of amino acids by collecting those
it encounters in the solution.  

Second, the membrane provides a medium for energy transfer
in which chemical potential energy couples to phonons in the surface.
This can assist the formation of bonds between amino acids,
and so overcome the barrier necessary to exclude water.

\end{itemize}

If these hypotheses are correct than the surfaces of vesicles will
accumulate a surface density of amino acids and short, random
peptides, that increase linearly for the time the vesicle remains
stable.  As vesicles have been shown to be stable over periods
of days\cite{bilayer} significant 
concentrations of polypeptides may
accumulate on their surfaces.

Here we may note an observation of 
Morowitz\cite{harold2}, which is
that short polypeptides bound to lipid membranes may be able
to catalyze reactions that would require longer proteins
in solution.  The reason is first, that the
membrane may provide structural support for the active
areas of the enzyme that in solution
is provided by the longer protein.  Second, the membrane may,
as already noted, play a role transferring energy, that in solution
would require a longer protein.

It follows from this observation that autocatalytic networks
of polypeptides as envisioned by Kauffman\cite{stu-origin} may be
more likely to develop attached to lipid-water boundaries
than in solution. 

Finally, we may observe that the formation of polypeptides
on the surface of a membrane may provide a mechanism
for breaking chirality.  By itself, the membrane, or course,
does not break the chiral symmetry.  But if a polypeptide
is to form such that all of its hydrophobic groups are
buried in the membrane, while all of its hydrophilic
groups stick out, and if the chain is to have bend
uniformly to form a circle or spiral on the surface,
then chains of amino acids which are dominantly of
a single chirality may be preferred energetically.

\subsection{What the polypeptides may do for the vesicles}

Even if the existence of proteins bound to the surface
or embedded within the membrane of the lipid vesicle leads to
the catalysis of metabolic reactions, there is still a problem
which must be overcome, which is the osmotic paradox.
How this is overcome is the subject of the next hypothesis.

\begin{itemize}

\item{}{\bf H3}  {\it There is a small, but non-vanishing probability
for a short random polypeptide made of random sequences
of amino acids that bind to membranes to bury themselves in
the membrane, and thus form channels through the membrane.}
When this occurs, the osmotic crisis is solved, for the channel
opened up by the polypeptide can allow the ions or molecules that
are the source of the osmotic pressure to escape.  This lowers
the osmotic pressure, and so ensures that the primitive metabolic
processes that are the source of the pressure do not destroy the
vesicle that contains them.  

\end{itemize}

This hypothesis may be restated the following way: {\it The first
biologically active proteins were primitive membrane channels.}
This closes the circle, the lipid membranes catalyze the synthesis
of short, random polypeptides from amino acids, while some
of the polypeptides so produced will serve as channels to 
release the pressure due to primitive metabolic processes.

There are several arguments which we believe support this
hypothesis beyond the need to resolve the osmotic paradox,
if vesicles which encapsulate primitive metabolic processes
are to be stable.

First of all, it is likely that an exact shape or sequence of
amino acids is not needed to form a primitive channel
in the membrane.  Any polypeptide formed on the surface
of a lipid vesicle as described above will be made of 
amino acids that are either hydrophobic, or contain
both hydrophobic and hydrophilic groups.  These naturally
want to bury pieces of themselves in the membrane.  All that
is needed is for the minimal energy state of such a polypeptide
in the surface to be a circle or spiral of sufficient length for
a channel to be drilled.  As shown by the properties of
antibiotics such as gramicidin, on the order of 10
amino acids is sufficient for a polypeptide to be a channel
former.

In contrast, a good deal more specificity is required for a 
polypeptide that catalyze a chemical reaction.  For this
reason it makes sense to hypothesize that protein channels
formed prior to enzymes.  

If this last hypothesis is true than we would expect vesicles
on which peptides had assembled themselves into channels to
be more stable, and hence last longer, than those that do not.
As we have already noted, there is a natural mechanism for 
vesicles to reproduce themselves by accumulation of lipids
followed at some point by spontaneous division.  To have
evolution of the vesicles and protein channels there must
be also a process of self-reproduction of the peptides that
form the primitive channels.  We may note that processes in
which short peptides catalyze their own reproduction have
recently been observed\cite{reza1}.  We need then the final
hypothesis.

\begin{itemize}

\item{}{\bf H4} {\it There is a non-vanishing probability for
an autocatalytic network of polypeptides to develop on the
surface of a vesicle, one of whose reactions would lead 
to the reproduction of polypeptides
which form membrane channels.}  When this occurs, that autocatalytic
network, together with the vesicle on which it grows, will have
the ability of reproducing itself, when the vesicle divides.

\end{itemize}

If these hypotheses are true one would expect to find somewhere
in biology short polypeptides with the following two properties:
1)  when they come into contact with a lipid membrane they
burrow into it and drill a channel through it and 2) they are
synthesized through some process which is not the standard
template process.  The reason is that, given
the hypothesis we have made here, there must have
evolved primitive mechanisms of synthesis and control
of proteins in the period between the
one described here and the eventual takeover of control of 
the mechanisms of the cell by nucleic acids.  As
these were prior to the nucleic acids and the formation of the
machinery of protein synthesis from amino acid templates, the
original method of synthesis must have involved a non-standard
series of reactions.  Given the principle that increases of 
complexity in biology almost always come from adding structure
and processes, while preserving the original 
system\cite{harold-sfi}, we would expect some of these
pre-nucleic acid proteins to be still active somewhere in biology and
to be synthesized by some non-standard, non-template process.
If protein channels were among the pre-nucleic acid proteins,
some polypeptides able to drill holes in membranes may have
survived which are still non-template synthesized.

Remarkably, at least one class of polypeptides that fulfills exactly
these functions does exist, it is the gramicidins\cite{gramicidin1}.
These are a class
of low molecular weight channel formers that are formed from
15 monomers.  Remarkably, they are made of both L and D
amino acids, which provides further evidence for their
antiquity.

\section{Conclusions}

Several key questions remain if these rough ideas are to
provide a useful starting point for experimental and
theoretical study.  

1)  Might there be a mechanism for  
the protein channels reproduce themselves, using only the
lipid membranes and proteins?   We know that short
peptides on the order of 32 amino acids are able to
spontaneously reproduce themselves from shorter 
elements\cite{reza1}.
It would be very interesting to discover a short polypeptide that
both forms channels in lipid membranes and 
catalyzes its own synthesis out of shorter polypeptide.
Alternatively, what is required is only the discovery of
a short protein channel which is part of an autocatalytic
set of polypeptides which may be bound to the surface
of a lipid bilayer.  We may note that $100 \%$ faithful
reproduction is not at all required, it would be sufficient
to have something like a quasispecies of polypeptides that
formed autocatalytic sets that reproduce themselves with
sufficient fidelity to be self-sustaining.

2)  We have kept a part of the story in the background, due
to our ignorance.  This is the form of the primitive metabolism
that we hypothesize gets started in the interiors of the lipid
vesicles\cite{harold2,harold-sfi}, whose 
action sets up the osmotic crisis.  Presumably
to become self-sustaining, the system of primitive protein
channels must become coupled to the primitive metabolism.   
There are natural ways for this to happen, as the products of
the metabolism in fact pass through the protein channels.

This is in fact
necessary as the channels must have some specificity;  their effect
must be not just to open holes in the membranes but to control
the passage of specific ions and molecules through it.
It may then be natural for channels to evolve that gain
energy from the osmotic pressure of the passage of ions or
molecules through them, and then contribute this energy
to the autocatalytic cycles that  lead to their own reproduction.
It would be important to identify possible mechanisms for this
coupling.  

3)  One way to test these ideas would be to study the interaction
of gramicidins and other channel formers with possibly primitive
membranes such as those that form spontaneously in 
experiments.

4)  It would also be of interest to catalogue and study those
biological polypeptides and proteins that are either not
template synthesized or include amino acids of the wrong
chirality.  These are candidates for primitive proteins from
the conjectured pre-nucleic acid era.  On the other side, those
who hypothesize an RNA first scenario for the origin of life must
supply a reasonable explanation for the occurrence of biological
polypeptides that are not template synthesized.

5)  A slightly modified form of this proposal would regard the
vesicles and the chemical reactions they encapsulate as 
initially part of the
environment and the membrane channels themselves as the
first autonomous agents, or self-reproducing entities.
This picture would be relevant if there were a short polypeptide
that reproduced itself, following something like the mechanism
of \cite{reza1} using energy gained by the release of ions 
(produced by the chemical reactions in the vesicle) through
a channel that it opened up in the vesicle's membrane.   
In the event that the
new protein channel was released into solution there might 
first evolve
a population of channel openers in solution that reproduced
themselves by fastening onto a primitive vesicle and drilling
a channel in it in order to gain the energy needed to catalyze
the linkage of smaller polypeptides into new copies of themselves.

The cell would then form when a population of such channel
openers evolved that could control the energy released sufficient
to extend the lifetime of the vesicle.  This would open the way
to a coevolution of the vesicles and channel openers.

\section*{ACKNOWLEDGEMENTS}

We are grateful especially to Stuart Kauffman and Harold
Morowitz for many conversations about the origin of life
and this particular proposal.  We thank them and David
Deamer and Andy Ellington also for very helpful comments
on a draft of this paper.  We are also grateful to
Federico Moran and Peter Schuster for lectures on the
origin of life which were helpful in formulating this proposal.
Conversations with Mark Goulian, Albert Libchaber
and Marcelo Magnasco about gramicidin
and the origin of life in general were most helpful, and these
were made possibility by the hospitality of Prof. Nick Khuri
at Rockefeller University.
This work was 
supported by a NASA grant to The Santa Fe Institute.

\end{document}